\documentclass[
 preprint,
 superscriptaddress,
 amsmath,amssymb,
 aps,
 prl,
 longbibliography,
 showpacs
]{revtex4-1}
\usepackage{graphicx}
\usepackage{dcolumn}
\usepackage{bm}
\usepackage{epstopdf}
\begin{document}

\title{Enhanced emission from a single quantum dot in a microdisk at a deterministic diabolical point}
\author{Jingnan Yang}

\author{Shushu Shi}
\author{Xin Xie}
\author{Shiyao Wu}
\author{Shan Xiao}
\author{Feilong Song}

\author{Jianchen Dang}
\author{Sibai Sun}
\author{Longlong Yang}

\affiliation{Beijing National Laboratory for Condensed Matter Physics, Institute of Physics, Chinese Academy of Sciences, Beijing 100190, China}
\affiliation{CAS Center for Excellence in Topological Quantum Computation and School of Physical Sciences, University of Chinese Academy of Sciences, Beijing 100049, China}
\author{Yunuan Wang}
\affiliation{Beijing National Laboratory for Condensed Matter Physics, Institute of Physics, Chinese Academy of Sciences, Beijing 100190, China}
\affiliation{Key Laboratory of Luminescence and Optical Information, Ministry of Education, Beijing Jiaotong University, Beijing 100044, China}
\author{Zi-Yong Ge}
\author{Bei-Bei Li}

\author{Zhanchun Zuo}
\affiliation{Beijing National Laboratory for Condensed Matter Physics, Institute of Physics, Chinese Academy of Sciences, Beijing 100190, China}
\affiliation{CAS Center for Excellence in Topological Quantum Computation and School of Physical Sciences, University of Chinese Academy of Sciences, Beijing 100049, China}
\author{Kuijuan Jin}
\affiliation{Beijing National Laboratory for Condensed Matter Physics, Institute of Physics, Chinese Academy of Sciences, Beijing 100190, China}
\affiliation{CAS Center for Excellence in Topological Quantum Computation and School of Physical Sciences, University of Chinese Academy of Sciences, Beijing 100049, China}
\affiliation{Songshan Lake Materials Laboratory, Dongguan, Guangdong 523808, China}
\author{Xiulai Xu}
\email{xlxu@iphy.ac.cn}
\affiliation{Beijing National Laboratory for Condensed Matter Physics, Institute of Physics, Chinese Academy of Sciences, Beijing 100190, China}
\affiliation{CAS Center for Excellence in Topological Quantum Computation and School of Physical Sciences, University of Chinese Academy of Sciences, Beijing 100049, China}
\affiliation{Songshan Lake Materials Laboratory, Dongguan, Guangdong 523808, China}



\begin{abstract}
We report on controllable cavity modes through controlling the backscattering by two identical scatterers. Periodic changes of the backscattering coupling between two degenerate cavity modes are observed with the angle between two scatterers and elucidated by a theoretical model using two-mode approximation and numerical simulations. The periodically appearing single-peak cavity modes indicate mode degeneracy at diabolical points. Then interactions between single quantum dots and cavity modes are investigated. Enhanced emission of a quantum dot with a six-fold intensity increase is obtained in a microdisk at a diabolical point. This method to control cavity modes allows large-scale integration, high reproducibility and flexible design of the size, location, quantity and shape for scatterers, which can be applied for integrated photonic structures with scatterer-modified light-matter interaction.
\end{abstract}
\maketitle

\section{Introduction}
Optical whispering-gallery microcavities with high quality (Q) factors and small mode volumes can enhance the interactions between light and matter\cite{Vahala2003}, and have found widespread applications in nonlinear optics\cite{Kuo2014}, low-threshold lasers\cite{Mao:11,OEA_RefItem:1}, dynamic filters and switches\cite{djordjev2002microdisk}, single-photon sources\cite{Wang2019}, and cavity quantum electrodynamics (CQED)\cite{https://doi.org/10.1002/lpor.201900425, Srinivasan2007,PhysRevLett.120.213901}. In a whispering-gallery microcavity, the counter-propagating degenerate eigenmodes with the same polarization and resonance frequency correspond to a diabolical point (DP) in the presence of mirror symmetry. However, intrinsic scatterers in microcavities will lift the mode degeneracy and cause the symmetry breaking. Embedded quantum dots (QDs)\cite{Jones:10,Hiremath:08}, material inhomogeneity\cite{Weiss:95,Kippenberg:02}, attached particles or fabrication imperfections\cite{He_2013,Li:12,Yang2020} can serve as intrinsic Rayleigh scatterers in microcavities and cause backscattering of the light into the counter-propagating mode and couple two counter-propagating modes with each other, lifting the mode degeneracy. Two new standing-wave modes will form and split in frequency, propagating with a phase $\pi/2$ between their spatial field distributions. The backscattering from the uncontrollable intrinsic scatterers will bring unwanted disturbances to the applications which in principle use the counter-propagating degenerate modes, such as particle sensors\cite{Zhu2009, Armani783}, dual frequency microcombs\cite{Yang2017}, and optical gyroscopes\cite{Liang:17}.

  Moreover, the backscattering can be precisely controlled by coupling nanotips to a microcavity \cite{article3, article4, Svela2020}, which have been used to realize exceptional points for chiral lasing\cite{PMID:27274059}, high-sensitivity particle detection\cite{Chen2017} and electromagnetically induced transparency\cite{Wang2020}. However, the above mentioned controlling method requires high-precision piezo positioners, which is challenging for large-scale integration. Instead, by introducing two subwavelength-scale perturbations as scatterers on the microcavity edge and adjusting the size, location, quantity and shape of the perturbations, backscattering can be controlled in an integrable way\cite{Kim:14}. Introduced designed scatterers have been used for lasing mode selection\cite{Schlehahn:13} and unidirectional lasing\cite{Wang22407} by tuning the wave propagations and mode field distributions in microcavities.

  QDs, also called the artificial atoms, have great potential for realizing ideal single-photon or entangled-photon sources\cite{doi:10.1063/1.2437727, Michler2282, Huber2017,doi:10.1063/1.3522655,Lohrmann2017}. To improve the indistinguishability of photons and enhance the photon extraction efficiency\cite{Liu2018,Kimble_1998,Wang2019}, QDs are usually weakly coupled to a cavity with high Q factors. Since spontaneous emission arises due to interactions between matter and its local electromagnetic environment\cite{Purcell1995, Pelton2015}, the interactions between a single QD and cavity modes depend on their spatial and frequency overlaps. The randomly-distributed intrinsic scatterers can cause the mode field redistribution of two newly formed modes in space and frequency including the mode field distribution surrounding QDs\cite{PhysRevA.75.023814}. But these intrinsic scatterers affect the cavity modes in an uncontronllable way. In addition, split modes will bring unwanted mode competition and prevent the single-mode lasing in a microcavity\cite{Zhu:18,Kim2016} with embedded QDs as gain medium for ultra-low threshold lasers\cite{Zhou:19,Zhang2020}. Therefore, to realize controlled backscattering is highly desired to control the cavity modes in microcavities with embedded QDs.

  Here we realize controllable cavity modes through controlling backscattering in GaAs microdisks with embedded QDs. We design microdisks with two identical nanoscale cuts serving as Rayleigh scatterers on the cavity perimeter and control the backscattering through changing the relative angle between two cuts. Periodic changes of the backscattering coupling characteristics of cavity modes in fabricated microdisks are obtained. A theoretical model based on two-mode approximation and three-dimensional numerical simulations are used to explain the periodic behaviours\cite{PhysRevA.84.063828}, in particular, the periodic mode degeneracy and difference in frequency and linewidth of split modes. Split cavity modes periodically merge into single-peak cavity modes, indicating DPs. Multiple cavity modes are also simultaneously controlled by scatterers in a periodical way. To investigate the interactions between single QDs and cavity modes, we calculate optimal Purcell factors and experimentally obtain enhanced photoluminescence (PL) intensity of QDs. We obtain enhanced PL intensity of a single QD with an approximately six-fold intensity increase in a microdisk at a deterministic DP.

\section{Theory}
For an unperturbed whispering-gallery microcavity, light can be confined as degenerate travelling-wave modes propagating along clockwise (CW) and counter-clockwise (CCW) directions, which correspond to two orthogonal states with coalescent eigenvalues. This indicates that the microcavity is at a DP. The Hamiltonian of an unperturbed microcavity in the travelling-wave basis can be written as
\begin{equation} \label{h0}
H_0 = \left(\begin{array}{cc}
\Omega-i\Gamma/2 & 0 \\
0 & \Omega-i\Gamma/2
\end{array}\right),
\end{equation}

where $\Omega-i\Gamma/2$ is the eigenvalues of $H_0$, $\Omega$ is the resonance frequency and $\Gamma/2$ is the decay rate. When adding two identical scatterers with size much smaller than the mode wavelength, backscattering of the mode fields from the scatterers will result in the coupling between the two degenerate modes and then two new modes are formed as standing-wave modes. The effective Hamiltonian of the whole system using the two-mode approximation can be represented by\cite{PhysRevA.84.063828}
\begin{equation} \label{h1}
H= H_0 + H_1,
\end{equation}
with
\begin{equation}
H_1 = \left(\begin{array}{cc}
\omega_{'} &  A \\
B & \omega_{'}
\end{array}\right),
\end{equation}
where $\omega^{'}\simeq 2\epsilon$, $A = \epsilon +\epsilon e^{-i2m\beta} $, and $B = \epsilon +\epsilon e^{i2m\beta}$. $m$ is the azimuthal mode number, $\beta$ is the angle between two scatterers on a microcavity, and $\epsilon$ is a complex frequency splitting resulting from the perturbation by a scatterer, defined as $\epsilon=g-i\Gamma_{1}/2$\cite{Zhu2009}. $g$ is the resonance frequency shift and $\Gamma_{1}/2$ is the additional decay rate caused by the scatterer. The offdiagonal elements $A$ and $B$ represent two light backscattering processes from CW to CCW and CCW to CW propagation directions, respectively. Obviously, $A$ and $B$ have periodic components with a calculated period of $P_{cal}=\pi/m$. Here $|A|=|B|$ indicates the symmetric backscattering and no chirality for the cavity\cite{PMID:27274059}. Therefore, we have the total effective Hamiltonian in the travelling-wave basis as
\begin{equation}
H = \left(\begin{array}{cc}
\Omega-i\Gamma/2+2\epsilon &  A \\
B & \Omega-i\Gamma/2+2\epsilon
\end{array}\right),
\end{equation}
and the new eigenvalues of this cavity are
\begin{equation} \label{h2}
\omega_{\pm}=\Omega-i\Gamma/2+2\epsilon \pm 2\epsilon \cos(m\beta).
\end{equation}

   To analyse the effects from two scatterers on the cavity, three parameters describing the backscattering coupling characteristics of the newly formed modes can be defined as
\begin{equation} \label{h3}
\Delta \omega = 4g\cos(m\beta),
\end{equation}
\begin{equation} \label{h4}
\Delta\lambda_{diff}=2\Gamma_{1}\cos(m\beta),
\end{equation}
\begin{equation} \label{h5}
Q_{sp}=\dfrac{2g\cos(m\beta)}{\Gamma/2+ \Gamma_{1}}.
\end{equation}
Here $\Delta \omega=Re(\omega_{+})-Re(\omega_{-})$ is the frequency splitting between two cavity modes. $\Delta\lambda_{diff}=-2[Im(\omega_{+})-Im(\omega_{-})]$ is the linewidth difference between two split modes, indicating the energy decay rate difference between two split modes. $Q_{sp}=2\Delta \omega /\Delta\lambda_{sum}$ is the splitting quality factor with $\Delta\lambda_{sum}=-2[Im(\omega_{+})+Im(\omega_{-})]$\cite{PMID:27274059}. When $Q_{sp}>1$, two split modes can be observed in spectra, otherwise a mode broadening or a frequency shift can be observed\cite{PhysRevA.97.063828}. Obviously, $\Delta \omega$, $\Delta\lambda_{diff}$ and $Q_{sp}$ are all periodically controlled by $\cos(m\beta)$.

    When $\beta=(2N+1)\pi/2m$ ($N$ is an integer), for example, for $N=9$ and $m=10$, then $\cos(m\beta)=0$ with $\beta=171.0^{\circ}$ is obtained. In this case, $A= B = 0$ is satisfied and the two eigenvalues $\omega_{\pm}$ coalesce with two new orthogonal modes, corresponding to a DP. $A = 0$ ($B = 0$) means the backscattering of CW (CCW) propagating wave at two scatterers is interfering destructively. The new coalescent eigenvalues $\omega_{\pm}=\Omega-i\Gamma/2+2\epsilon$ contain an additional energy decay rate $\Gamma_{1}$ and a resonance frequency shift $2g$ from $2\epsilon$. Furthermore, when $\beta=N\pi/ m$, for example, for $N=10$ and $m=10$, then $\omega_{\pm}=\Omega-i\Gamma/2+2\epsilon \pm 2\epsilon$ with $\beta=180.0^{\circ}$. $\Delta \omega$, $\Delta\lambda_{diff}$ and $Q_{sp}$ will all reach their maximum values.

   The linewidth distributions of the new modes with the corresponding relative wavelengths can be specifically inferred according to Equation (\ref{h3}) and (\ref{h4}). The sign of $g$ depends on the dielectric permittivity difference $\varepsilon_{m}-\varepsilon_{p}$ between the surrounding media ($\varepsilon_{m}$) and the scatterers ($\varepsilon_{p}$)\cite{Zhu2009}. When $\varepsilon_{m}< \varepsilon_{p}$, then $g<0$. $\Delta \omega$ and $\Delta\lambda_{diff}$ will exhibit opposite signs, which means the new mode with a longer wavelength will exhibit a wider linewidth. When $\varepsilon_{m}> \varepsilon_{p}$, then $g>0$. $\Delta \omega$ and $\Delta\lambda_{diff}$ will exhibit the same signs, which means the new mode with a shorter wavelength will exhibit a wider linewidth. Here we assume $\varepsilon_{m}> \varepsilon_{p}$ since the effective dielectric permittivity $\varepsilon_{p}$ of the designed scatterers in our devices is 1. Therefore, for $\beta=N\pi/ m$, one of the new eigenvalue will exhibit a smaller imaginary part than that of the above coalescent eigenvalues and corresponds to a longer resonance wavelength as shown in Fig. \ref{p5}(a) and (b).

   Additionally, when $A=0$, $B\neq 0$ or $B=0$, $A\neq 0$ is satisfied, exceptional points appear where both eigenvalues and eigenvectors of the system coalesce with asymmetric backscattering\cite{PMID:27274059}. Different from the identical scatterers to achieve a DP with symmetric backscattering, the two scatterers to achieve an exceptional point should have different sizes and distances from cavity rim to break the mirror symmetry and achieve a high asymmetry in backscattering at $\beta=(2N+1)\pi/2m$\cite{PhysRevA.84.063828}.

\section{Design and methods}
\begin{figure}
\centering
\includegraphics[scale=0.5]{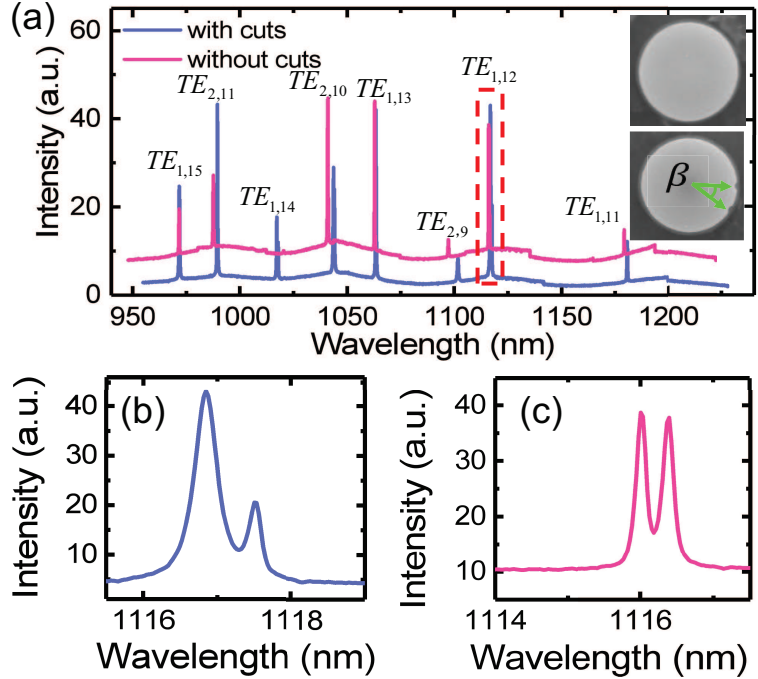}
\caption{(a) PL spectra of cavity modes from microdisks with (blue line, $\beta = 50^{\circ}$) and without cuts (pink line) under a high excitation power. The right insets: scanning electron microscope (SEM) images of microdisks with (bottom) and without (top) cuts. The cuts with $\beta$ between them are marked out by the green arrows. (b)-(c) Enlarged PL spectra of split modes from (a) in the red dashed rectangle for microdisk with (b) and without (c) cuts, respectively.}
\label{p1}
\end{figure}

  The sample grown by molecular beam epitaxy is made of three layers, namely GaAs substrate, a 1-$\mu m$-thick AlGaAs sacrifice layer and a 250-$nm$-thick GaAs slab with InAs QDs grown in the middle. To investigate the interactions between single QDs and cavity, the QDs are intentionally grown in an appropriately low density of approximately 30 $\mu m^{-2}$ for distinguishing the discrete spectral emissions of single QDs. In order to investigate how the backscattering is controlled by the angle $\beta$ between two scatterers on the microdisk perimeter and guarantee the cavity mirror symmetry to achieve DPs\cite{Kim:14, PhysRevA.84.063828}, we design two identical arc cuts to serve as the scatterers through removing two circles off the cavity with the circle centers are on the microdisk perimeter as shown in the inset in Fig. \ref{p1}(a) and in Fig. \ref{p4}. $\beta$ is designed from 0$^{\circ}$ to 180$^{\circ}$ with 5$^{\circ}$ as an angle step. To check the device reproducibility, five identical microdisks for each parameter are fabricated. The microdisk radii are 1 $\mu m$. Cut radii $R_{cut}$ are ranging from 30 to 60 $nm$, which are much smaller than the cavity mode wavelength of approximately 300 $nm$ inside GaAs cavity (the refractive index of GaAs $n_{m}$ is 3.46). The fabrication process involves first structure patterning of the masks on the sample surface by electron beam lithography, followed by dry etching of the GaAs slab and the AlGaAs layer by inductively coupled plasma to form circular pillars, and the final wet etching of the AlGaAs layer by HF solutions to form a pedestal under each microdisk. The device is mounted on a nano-positioner and cooled down to 4.2 $K$ through the heat exchange between the device chamber and the liquid Helium bath with Helium gas. The device temperature can be changed by a heater mounted on the nano-positioner. A magnetic field can be applied by superconducting coils in the chamber to tune the quantum dot emission energy. A laser with a wavelength of 532 $nm$ in a conventional confocal micro-PL system is used to excite the device. The PL spectra are collected by a linear array of InGaAs detectors after a monochromator.
\section{Periodic backscattering in microdisks}
  Figure \ref{p1}(a) shows PL spectra of cavity modes exicted by PL from ensemble of QDs in microdisks with (blue line) and without (pink line) cuts, where $R_{cut}$= 40 $nm$ and $\beta = 50^{\circ}$. The cavity modes are respectively distinguished with the corresponding mode numbers according to three-dimensional simulation results using finite element method. Here only more localized transverse electric (TE) polarized modes in the first and the second radial order are observed in such a thin disk with a thickness of 250 $nm$\cite{Srinivasan:06}. The cavity modes are notated by $TE_{p,m}$, where $p$ is the the radial mode number and $m$ is the azimuthal mode number. Insets in Fig. \ref{p1}(a) are SEM images of fabricated microdisks with (down) and without (top) cuts. The enlarged PL spectra of modes in dashed rectangular are shown in Fig. \ref{p1}(b) and (c). The split modes from the microdisk with cuts in Fig. \ref{p1}(b) exhibit two unequal linewidths according to Equation (\ref{h4}). That the left branch of the double peaks exhibits a wider linewidth than that of the right branch is due to a smaller effective dielectric permittivity of the scatterers compared to that of the microdisk\cite{Zhu2009}, as discussed above in $Theory$ section. Due to the intrinsic scatterers, split modes exhibiting randomness were also observed in the microdisks without cuts as shown in Fig. \ref{p1}(c). By comparison, the split peaks exhibit two almost equal linewidths resulted from a large amount of randomly distributed scatterers\cite{Yang2020}.

\begin{figure}
\centering
\includegraphics[scale=0.5]{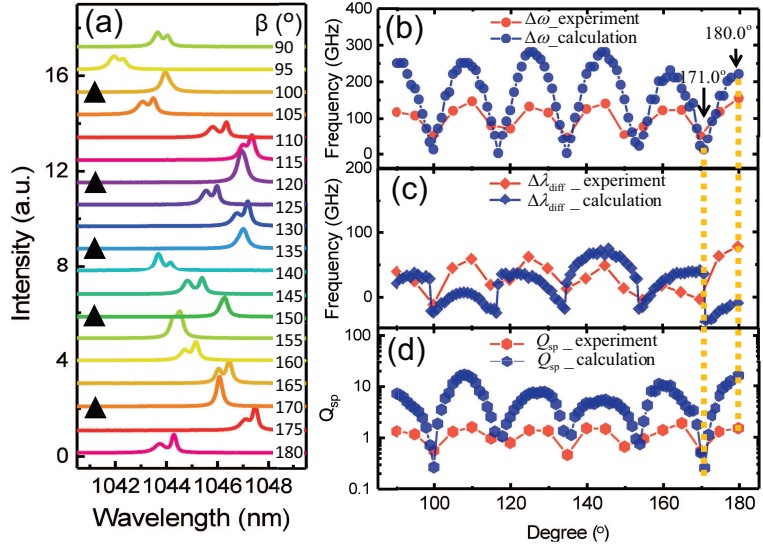}
\caption{(a) PL spectra of cavity mode $TE_{2,10}$ from microdisks with two cuts where $\beta$ is from $90^{\circ}$ to $180^{\circ}$ and $R_{cut}$=40 $nm$. Single peaks, corresponding to the reappearing DPs, are marked with black triangles. (b)-(d) Average experimental (red) fitted results and simulation results (blue) of $\Delta \omega$ (b), $\Delta\lambda_{diff}$ (c) and $Q_{sp}$ (d) with $\beta$ from $90^{\circ}$ to $180^{\circ}$ and $R_{cut}$=40 $nm$. Experimental and simulated $\Delta \omega$, $\Delta\lambda_{diff}$ and $Q_{sp}$ all exhibit almost the equal and synchronous periods of about $18^{\circ}$. $171^{\circ}$ and $180^{\circ}$ correspond to $N=9$ and $m=10$ for $\beta=(2N+1)\pi/2m$ and $N=10$ and $m=10$ for $\beta=N\pi/ m$, respectively.}
\label{p2}
\end{figure}

  Compared with the random and uncontrollable intrinsic backscattering, periodically controlled backscattering coupling characteristics including $\Delta \omega$, $\Delta \lambda_{diff}$ and $Q_{sp}$ are achieved by two designed cuts. Taking the mode with wavelength of approximately 1043 $nm$ ($TE_{2,10}$) as an example in microdisks with $R_{cut}$=40 $nm$ and $\beta$ from $90^{\circ}$ to $180^{\circ}$ with $5^{\circ}$ as one angle step, split cavity modes periodically merge into single-peak cavity modes marked by black triangles, as shown in Fig. \ref{p2}(a). The average fitted experimental results (red) from identically designed microdisks including $\Delta \omega$, $\Delta \lambda_{diff}$ and $Q_{sp}$ are shown in Fig. \ref{p2}(b), (c) and (d), respectively. Synchronously periodic changes of experimental $\Delta \omega$, $\Delta\lambda_{diff}$ and $Q_{sp}$ can be observed where they reach their extreme values at the same $\beta$, agreeing with the above conclusion that the backscattering characteristics are periodically controlled by $cos(m\beta)$. Therefore, those single-peak cavity modes in Fig. \ref{p2}(a) correspond to the reappearing DPs. Such periodic behaviours also happen to other modes in Fig. \ref{p1}(a) because the two scatterers affect all the cavity modes. Their experimental periods $P_{exp}$ are listed in Table \ref{t1}. Considering the angle step of $5^{\circ}$, errors for $P_{exp}$ of different modes are estimated by $\pm 10^{\circ}/N_{P}$. $N_{P}$ is the number of periods counted from 0 to 180$^{\circ}$.

\begin{table}[htbp]
\centering
\caption{\bf Simulated ($\lambda_{sim}$) and experimental ($\lambda_{exp}$) mode wavelengths in a microdisk with radius = 1 $\mu m$ and thickness of 250 $nm$ without cuts, and calculated ($P_{cal}$) and experimental ($P_{exp}$) periods in microdisks with radius = 1 $\mu m$ and $R_{cut}=$ 40 $nm$.}
\resizebox{300pt}{50pt}{
\begin{tabular}{ccccccc}
\\
\hline
Mode&$TE_{1,15}$&$TE_{2,11}$&$TE_{1,14}$&$TE_{2,10}$&$TE_{1,13}$&$TE_{1,11}$\\
\hline
$\lambda_{sim}/nm$&979&1002&1015&1043&1054&1189\\

$\lambda_{exp}/nm$&973&989&1012&1043&1064&1181\\

$P_{cal}/^{\circ}$&12.0&16.4&12.9&18.0&15.0&16.4\\

$P_{exp}/^{\circ}$&12.1$\pm0.7$&16.0$\pm 1.0$&13.1$\pm 0.8$&17.8$\pm 1.1$&13.8$\pm0.8$&16.3$\pm 1.3$\\
\hline
\end{tabular}}
\label{t1}
\end{table}

  To further analyse the backscattering controlled by the scatterers, we have done three-dimensional numerical simulations based on finite element method on a microdisk with two cuts. Specific modes are confirmed by comparing mode wavelength $\lambda_{exp}$ in Fig. \ref{p1}(a) with simulated wavelength $\lambda_{sim}$ in Table \ref{t1}. For more relations between backscattering and mode field distribution, we focused on the mode $TE_{2,10}$. Synchronously periodic changes of $\Delta \omega$, $\Delta \lambda_{diff}$ and $Q_{sp}$ as a function of $\beta$ are obtained as shown in Fig. \ref{p2}(b) and (d) labelled by blue geometries which exhibit almost the same periodicity as the experimental results, respectively. In comparison with simulation results, a smaller experimental $Q_{sp}$ is due to a lower Q factor caused by fabrication imperfection. The experimental $\Delta \omega$ are also smaller than simulation results, which can be attributed to that the size of etched cuts on the microdisks is smaller than designed. Instead, the cut size will decrease from top to bottom at the cavity perimeter by inductively coupled plasma. The smaller etched cuts can cause smaller splitting as shown in Fig. \ref{p5}(d). The simulated period is equal to the calculated period $P_{cal}=180^{\circ}/10=18.0^{\circ}$. $P_{cal}$ for other $TE_{1,m}$ and $TE_{2,m}$ are listed in Table \ref{t1}. The deviations of experimental periods from calculated ones are attributed to the insufficiently small step angle of 5$^{\circ}$ for $\beta$, which limits the control of the backscattering.

\begin{figure}
\centering
\includegraphics[scale=0.5]{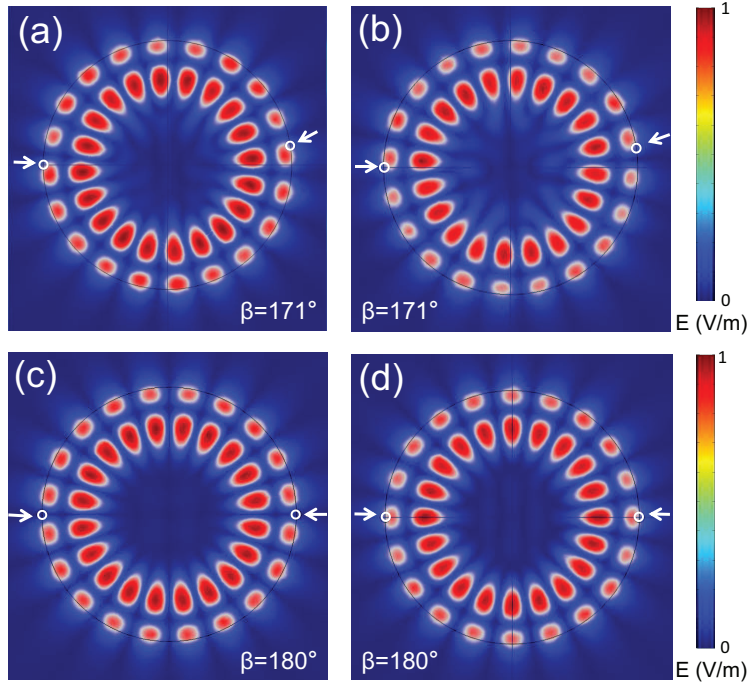}
\caption{Simulated electric field distribution of two pairs of modes $TE_{2,10}$ corresponding to two extreme values of the coupling backscattering characteristics. (a)-(b) Electric field distribution of two modes with $\beta=171.0^{\circ}$ corresponding to $N=9$ and $m=10$ for $\beta=(2N+1)\pi/2m$, where $\omega_{-} = 2.8766E14-i1.1905E9$ $Hz$ (a) and $\omega_{+} = 2.8767E14-i2.9221E9$ $Hz$ (b). Similar field distributions relative to cuts can be observed, indicating a DP with two degenerate modes. (c)-(d) Electric field distribution of split modes with $\beta=180.0^{\circ}$, where $\omega_{-} = 2.8753E14-i9.7229E9$ $Hz$ (c) and $\omega_{+} = 2.8775E14-i4.4371E9$ $Hz$ (d), corresponding to $N=10$ and $m=10$ for $m\beta=N\pi$. Inequivalent field distributions relative to cuts can be observed, where nodes in (c) and antinodes in (d) of field are at the cut center, respectively. The locations of cuts are marked with white arrows and the white circles indicate how the cuts are formed.}
\label{p4}
\end{figure}

   To visualize the mode degeneracy at DPs and how linewidths difference and frequency splitting of two modes are caused, simulated electric field of two pairs of cavity modes at $\beta = 171.0^{\circ}$ and $\beta = 180.0^{\circ}$ are shown in Fig. \ref{p4}, which are two opposite extreme cases for $\Delta \omega$, $\Delta\lambda_{diff}$ and $Q_{sp}$ as marked by yellow dashed lines in Fig. \ref{p2}. Figure \ref{p4}(a) and (b) show the mode field distributions for the pair of cavity modes at $\beta = 171.0^{\circ}$ corresponding to the minimum case in Fig. \ref{p2}. The field distributions of cavity mode near cuts in (a) and (b) are quite similar with antinodes deviating a little from the cut center in different directions, indicating the similar effects by the two scatterers. In this case, the two modes are degenerate at a DP, corresponding to $N=9$ and $m=10$ for $\beta=(2N+1)\pi/2m$. While Fig. \ref{p4}(c) and (d) show the pair of cavity modes at $\beta = 180.0^{\circ}$ corresponding to the maximum case in Fig. \ref{p2}. Two nodes of mode field distribution in Fig. \ref{p4}(c) are at the cut center, which means the cuts do not affect the cavity mode very much. Figure \ref{p4}(d) shows the opposite, where two antinodes of mode field are distributed at the cut center, indicating much stronger effects by cuts on energy loss and mode splitting. In this case, the effects from the scatterers on two mode fields are apparently unbalanced, resulting in the maximum values for $\Delta \omega$ and $\Delta\lambda_{diff}$ corresponding to $N=10$ and $m=10$ for $m\beta=N\pi$. Considering the mode field distributions, the cuts obviously cause the redistribution of the new mode fields, and the antinodes are controlled by the relative angle $\beta$. Therefore, by tuning the angle between scatterers, the mode field distribution surrounding a QD can be controlled for scatterer-modified light-matter interactions.

\section{Purcell enhanced emission from a single QD}
\begin{figure}
\centering
\includegraphics[scale=0.5]{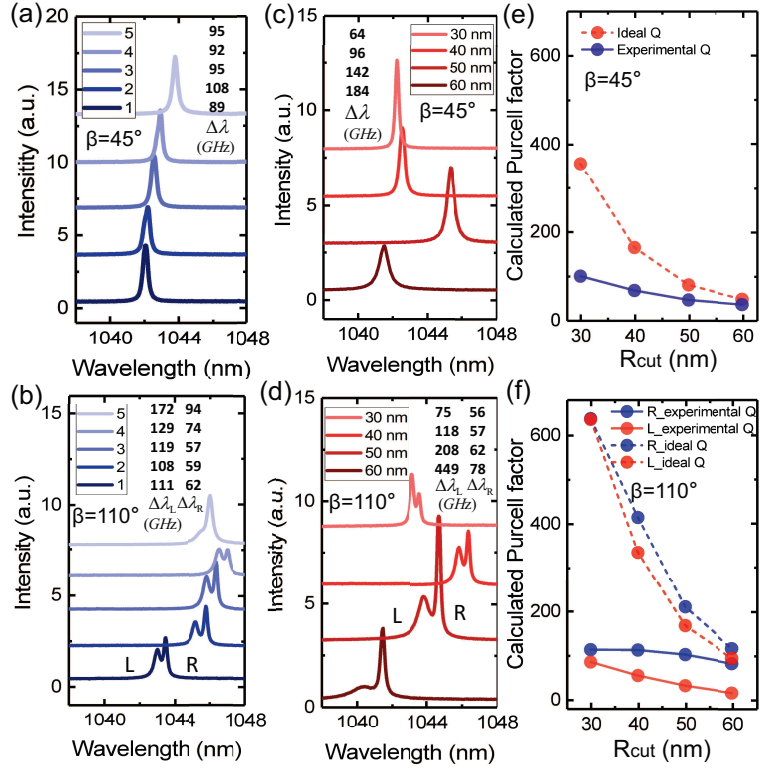}
\caption{(a)-(b) PL spectra of single peaks (a) and split peaks (b) for modes $TE_{2,10}$ from identically designed microdisks with $\beta=45^{\circ}$ and $\beta=110^{\circ}$, respectively. $\beta=45^{\circ}$ corresponds to $N=2$ and $m=10$ for $\beta=(2N+1)\pi/2m$ and $\beta=110^{\circ}$ approaches $\beta=N\pi/m=108^{\circ}$ ($N=6$ and $m=10$). The cut radii are 40 $nm$. (c)-(d) PL spectra of single-peak (c) and split-peak (d) modes from microdisks designed with different $R_{cut}$. (e)-(f) Calculated optimal Purcell factors for single-peak (e) modes from (c) and split-peak modes (f) from (d). The dashed line refers to calculated Purcell factors using ideal Q factors and the solid line refers to calculated Purcell factors using Q factors from the experimental spectra. $R$ and $L$ refer to the longer-wavelength and shorter-wavelength branches, respectively. The fitted linewidths of cavities modes and $\beta$ are presented in the insets.}
\label{p5}
\end{figure}

  To check the device reproducibility, identically designed microdisks are measured labelled by numbers from 1 to 5. Figure \ref{p5}(a) and (b) show modes $TE_{2,10}$ with $R_{cut}=$ $40 nm$ which correspond to single peaks at $\beta=45^{\circ}$ and split peaks at $\beta=110^{\circ}$, respectively. All the modes exhibit single peaks in Fig. \ref{p5}(a) with similar linewidth, corresponding to the mode degeneracy at DPs. Double peaks exhibit similar frequency splitting and linewidth distributions in Fig. \ref{p5}(b). The fitted linewidths of the cavity modes are shown in the insets accordingly. It can be seen that the right branches of the split modes exhibit smaller linewidths than those of the reappearing single-peak modes as theoretically predicted above. Hence, the spectral shapes of modes from identically designed devices are highly consistent, proving the high reproducibility of the devices. The minor wavelength differences are attributed to the fabrication imperfection.

  To study the effects of $R_{cut}$ on the backscattering, we measure microdisks with the same $\beta$ of $45^{\circ}$ and $110^{\circ}$ but different $R_{cut}$ (from 30 to 60 $nm$). Figure \ref{p5}(c) and (d) show that the increasing $R_{cut}$ barely affects the relative spectral shape but does lower Q factors. Obviously, the linewidths of left branches of the split modes increase with $R_{cut}$ while the linewidths of the right branches are barely affected, as exhibited by the fitted spectral linewidths in the insets. This is due to the unbalanced effects from the scatterers on two modes, as visualized in Fig. \ref{p4}(c) and (d). The frequency splitting between two split modes also increases with $R_{cut}$ because a bigger scatterer can cause more backscattering and further enhance the coupling between two degenerate modes\cite{He_2013}.

  Though the backscattering is well controlled by two cuts to modify cavity modes, it is at the cost of lowering Q factors. Therefore, to estimate the potentials of this control method, we calculate optimal Purcell factors in microdisks with different $R_{cut}$ and two different $\beta$ based on Q factors of our fabricated and ideal microdisks. Here we assume the QD is positioned at the maximum mode field of one cavity mode and the exciton dipole in the QD is parallel to the local mode electric field with much narrower linewidth than that of the cavity mode\cite{PhysRevA.97.043801}. Then the optimal Purcell factor is given by\cite{Vahala2003}

 \begin{equation} \label{h6}
F_{P}=\dfrac{3Q\lambda^{3}}{4\pi^{2}V n^{3}_{m}},
\end{equation}
 where $V$ is the mode volume and $n_{m}$ is the refractive index of cavity. For Purcell factors of the mode $TE_{2,10}$, we take $\lambda=1043$ $nm$, $n_{m} = 3.46$ and $V \approx 3.4\times\lambda^{3}/n^{3}_{m}$ calculated by finite element method. Figure \ref{p5}(e) and (f) show calculated Purcell factors of modes in Fig. \ref{p5}(c) and (d), respectively. Here each cavity mode is assumed to have a QD optimally positioned. The dashed lines refers to Purcell factors with ideal Q factors from simulations while the solid lines refer to those with experimental Q factors estimated by $Q=\lambda /\Delta \lambda$ from spectra in Fig. \ref{p5}(c) and (d). A sharp decrease of calculated Purcell factors with the increasing $R_{cut}$ can be observed due to the declining Q factors, indicating smaller cuts improving Q factors. The right branch of split modes shows a slower decline than the left branch.

  Through controlling the backscattering by two identical scatterers, we obtain microdisks at periodically deterministic DPs, where split cavity modes merge into single peaks. Then we focus on the interactions between single QDs and the cavity modes with backscattering control. At a high excitation power, PL of the ensemble of QDs can be observed in a wide spectrum range as shown in the background spectra of the cavity modes in Fig. \ref{p1}(a). To observe well-isolated discrete spectral lines from single QDs, a low excitation power is required. Due to the growth technology of QDs, QDs show randomness in its size and position\cite{RevModPhys.87.347} which means the QD may be blue-detuned or red-detuned from the cavity. For these two detuning cases, we use two methods by controlling the device temperature and a magnetic field to bring a single QD and a cavity mode into resonance in our experiment. When temperature increases, the emission energy of a QD redshifts, mainly ascribing to a shrinkage of the band gap of QDs\cite{PhysRevB.55.9757}. The energy of a cavity mode also redshifts but at a much slower speed than QDs due to an increase in the refractive index of the cavity\cite{Yoshie2004}. When a magnetic field is applied, the emission energy of a QD blueshifts due to diamagnetic effect while cavity modes are unaffected\cite{PhysRevB.66.193303,PhysRevB.57.9088,Cao2016,PhysRevLett.122.087401}. Therefore, for the QD blue-detuned from the cavity, we increase the temperature to tune the QD emission for resonance, while for the QD red-detuned from the cavity we apply a magnetic field to change the QD emission.

\begin{figure}
\centering
\includegraphics[scale=0.4]{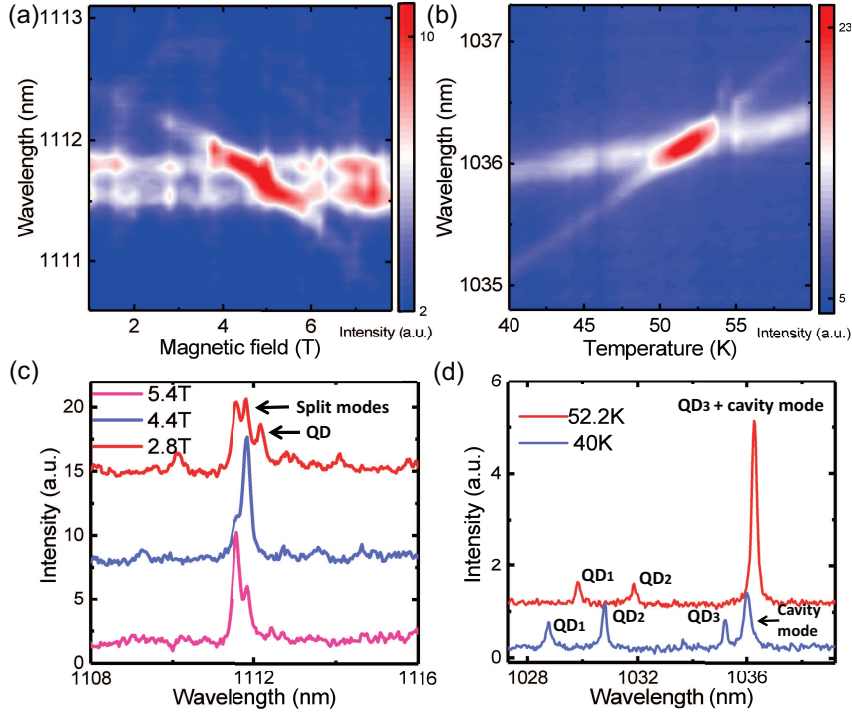}
\caption{(a) PL map for a single QD coupled to split modes $TE_{2,9}$ in a magnetic field. (b) PL map for a single QD coupled to single-peak mode at a DP with increasing the temperature. Purcell-enhanced emission from a single QD at resonance with cavity mode $TE_{2,10}$ at approximately 52.2 $K$. (c) PL pectra in a magnetic field of 2.8 $T$, 4.4 $T$ and 5.4 $T$ from (a) where the QD is off resonance with split modes, at resonance with the long-wavelength branch and the short-wavelength branch of split modes, respectively. (d) PL spectra at 52.2 $K$ and 40 $K$ from (b) which correspond to resonant and off-resonant cases, respectively. Three discrete spectral emission of three single QDs can be observed at 40 $K$ and red shifts of QD emission energies can be observed at 52.2 $K$.}
\label{p6}
\end{figure}

  Figure \ref{p6}(a) shows the coupling between a single QD and two split single modes where the spectral emission of the single QD blue shifts in an increasing magnetic field and then respectively crosses the two split modes indicating two weak couplings. As shown in Fig. \ref{p6} (c), the QD first couples with the long-wavelength branch of split modes and then with the short-wavelength branch where emission enhancements can be observed respectively at 4.4 $T$ and 5.4 $T$. Due to the spatial overlap of a QD with two split standing-wave modes is usually different, therefore enhancements are different depending on the location of the QD\cite{PhysRevA.75.023814}. Here we estimate the emission intensity of the QD coupled with two branches of split modes. For the right branch, the emission intensity is enhanced by approximately two times at 4.4 $T$ while approximately three times at 5.4 $T$ for the left branch.

  Furthermore, we investigate the interactions between a single QD and cavity at a DP. As shown in Fig. \ref{p6}(b), a single QD labelled by $QD_{3}$ in Fig. \ref{p6}(d) is tuned into resonance with a single-peak mode at a deterministic DP with $R_{cut}$=40 $nm$ by increasing the temperature. Both of the emission wavelengths of the single QD and the cavity mode red shift while the previous one shifts faster. A spectral crossing for the QD and the cavity mode can be observed, where two peaks merge into one peak with an obvious emission enhancement region and then separate with further increasing the temperature. This indicates weak coupling between the QD and the cavity mode. The merged peak reaches its maximum intensity at approximately 52.2 $K$. By comparing the off-resonant spectrum at 40 $K$ and the resonant spectrum at 52.2 $K$, Purcell-enhanced PL intensity of the single QD can be more apparently observed as shown in Fig. \ref{p6}(d). By Lorentz fitting the peaks, the PL intensity of the single QD is enhanced by approximately six times. It should be noted that the factor of the emission intensity enhancement is not sufficient to reveal the specific Purcell factor without lifetime measurement. The enhancement is strongly limited by the spatial overlap between the single QD and the mode field. Therefore, a small spatial overlap will result in a much weaker emission enhancement compared to the optimal Purcell enhancement where the QD is assumed optimally position in the mode field. Two discrete PL peaks from other two single QDs labelled by $QD_{1}$ and $QD_{2}$ which are uncoupled to cavity modes can also be observed in Fig. \ref{p6}(d) for comparison. With a temperature increase, emissions of two QDs shift to red as expected.
\section{Discussion}

  Previously, researchers have been focused only on the original mode degeneracy but with inevitable split cavity modes\cite{Jones:10,Srinivasan:06,Srinivasan2007} or backscattering control by nanotips requiring high-precision piezo positioners. Here, the backscattering is periodically controlled by the designed scatterers in microdisks with the changing angle $\beta$. Periodic mode degeneracy at DPs and Purcell-enhanced emission from a single QD coupled to a microdisk at a deterministic DP are obtained. Designed scatterers can control the mode field distribution,such as mode antinodes, indicating that whispering-gallery cavity with quantum emitters and controlled scatterers can provide a good platform to study scatterer-modified light-matter interactions. In addtion, designed scatterers can even be used to achieve exceptional points, which will extend studying light-matter interactions into a non-Hermitian system. The scatterers can be designed flexibly in size, location, quantity and shape. Smaller increasing angle steps for $\beta$ can also be designed for more precise controlled backscattering. Meanwhile, with further optimization of the fabrication, a cavity both with higher Q factors and with sufficiently controlled backscattering can be obtained by smaller scatterers. As the QD growth and imaging technology being improved \cite{Hennessy2007,Gschrey2015, doi:10.1063/1.4773882}, pre-selected and accurately aligned QDs could be beneficial to optimize the spatial overlap between QDs and controlled cavity mode field. Therefore, interactions between light and QDs with backscattering control can be greatly enhanced in the future.

\section{Conclusion}
  In summary, we demonstrate the control of cavity modes through controlling the backscattering by two identical designed scatterers integrated into optical microdisks. Periodic control of backscattering coupling characteristics of cavity modes are obtained when the angle between the two scatterers is changed. The periodically appearing single peaks indicate the periodic mode degeneracy at DPs. Theoretical discussion and numerical simulations are presented for understanding the periodic backscattering coupling characteristics. In particular, mode degeneracy, linewidth difference and frequency splitting at two extreme cases are discussed in theory part and visualized by the simulated mode field distributions. Meantime, multiple cavity modes are simultaneously controlled by scatterers but with different periods, which agree well with theory. To prove the potential application in light-matter interactions with cavity modes controlled by scatterers, optimal Purcell factors are calculated. Purcell-enhanced PL emission from a single QD with a six-fold intensity increase is experimentally obtained in a cavity at a deterministic DP. The enhancement can be further improved with increasing the spatial overlapping between QDs and cavity mode and the Q factors. We believe that whispering-gallery cavities with designed and integrated scatterers provide a good platform to study the light and matter interaction in a controllable way, which have great potentials for implementing optical quantum information processing and cavity quantum electrodynamics \cite{Chen2020, PhysRevLett.120.065301, Pick:17}.

\section{Acknowledgement}
This work was supported by the National Natural Science Foundation of China (Grants No. 62025507, No. 11934019, No.11721404 and No. 11874419), the Key-Area Research and Development Program of Guangdong Province (Grant No.2018B030329001), the Strategic Priority Research Program (Grant No. XDB28000000), the Instrument Developing Project (Grant No. YJKYYQ20180036) and the Interdisciplinary Innovation Team of the Chinese Academy of Sciences.

\section*{Disclosures}
The authors declare no conflicts of interest.

%





\end{document}